\documentclass[prl,aps,showpacs,twocolumn]{revtex4}
\date{ }
\usepackage{epsfig}

\begin{document}
 
\title{Fluctuations and stochastic processes in one-dimensional many-body quantum systems}

\author{H.-P. Stimming$^1$, N. J. Mauser$^1$,  J. Schmiedmayer$^2$ and I. E. Mazets$^{1,2,3}$}
\affiliation{$^1$Wolfgang Pauli Institute c/o Univ. Wien, Nordbergstrasse 15, 
1090 Vienna, Austria  \\
$^2$Atominstitut -- TU Wien, Stadionallee 2, 1020 Vienna, Austria \\
$^3$Ioffe Physico-Technical Institute, 194021 St.Petersburg, Russia}

\begin{abstract}

We study the fluctuation properties of a one-dimensional many-body quantum system composed of interacting bosons, and investigate the regimes where  quantum noise or, respectively, thermal excitations are dominant.  For the latter we develop a semiclassical description of the fluctuation properties based on the Ornstein-Uhlenbeck stochastic process. As an illustration, we analyze the phase correlation functions and the full statistical distributions of the interference between two one-dimensional systems, either independent or tunnel-coupled and compare with 
the Luttinger-liquid theory.

\end{abstract}

\date{\today}

\maketitle 

Measurement of fluctuations and their correlations yields important information 
on regimes and phases of many-body quantum systems \cite{altman04}. In ultracold atomic systems, 
these correlations revealed the Mott insulator phase of bosonic \cite{bloch1} and fermionic \cite{bloch2} 
atoms in optical lattices, they allowed detection of correlated atom pairs in
spontaneous four-wave mixing of two colliding Bose-Einstein condensates \cite{asp1} 
and Hanbury-Brown-Twiss correlation for non-degenerate metastable $^3$He and $^4$He atoms \cite{asp2} 
and in atom lasers \cite{esslinger}.  Furthermore, they have allowed studies of dephasing \cite{deph-s} 
and have been employed as noise thermometer \cite{gati06,tnoise-s}.

A key question in the physics of many-body quantum systems at finite temperature is how much of the observed fluctuations and their correlations are fundamentally quantum, and which are caused by the thermal excitations in the system.  In the present Letter we address this problem starting from a description of the excitations in the system and their occupation numbers. This allows us to directly explore the contributions of quantum (ground state) noise and thermal excitations.  
 
We consider a quantum degenerate spin-polarized gas of bosonic atoms in an extremely anisotropic trap, with transversal confinement frequency $\omega_\perp$ much larger then the longitudinal confinement frequency $ \omega_\parallel$ (typically ${\omega_\perp}/{\omega_\parallel} > 1000$). If both the temperature $T$ and the mean-field interaction energy per atom are small compared to the radial confinement  ($k_\mathrm{B}T \ll \hbar \omega_\perp$, $n_\mathrm{1D}a_\mathrm{s}\ll 1$, where $n_\mathrm{1D}$ is the linear atom-number density and $a_\mathrm{s}$ is the atomic $s$-wave scattering length), the atomic motion  is confined to the radial ground state of the trapping potential.  In this  1D regime a ``quasi-condensate" emerges, which can be characterized by a macroscopic wave function with a fluctuating phase \cite{Ha81,P00,Hann}.   

The statistical properties of the fluctuating phase are the focus of this study.  They can be probed by interfering two identically prepared 1D systems by creating quasi-condensates in two parallel, identical traps \cite{rfs}. When released they expand freely, overlap and interfere.  The local phase of the interference reflects the fluctuating relative phase $\theta (z)$ of the quasi-condensates. The fluctuations in $\theta (z)$ manifest themself in the phase correlation function $ {\cal C}_\theta (z-z^\prime) = \left\langle \exp [\mathrm{i}\theta(z)-\mathrm{i}\theta(z^\prime )]\right\rangle $. The full distribution function of the interference 
contrast has been derived in Refs. \cite{A1,A2}.

In our investigation we consider the general case of two 1D quasi-condensates which can be tunnel-coupled to each other, described by the effective Hamiltonian \cite{B2}  
\begin{eqnarray} 
\hat{\cal H}&=&\int dz\, { \{ }\sum _{j=1}^2\left[ \hat{\psi }^\dag _j(z)\hat{\cal T} 
\hat{\psi } _j(z) +\frac g2 \hat{\psi }^{\dag\, 2}_j(z)\hat{\psi }^2_j(z)\right] -\nonumber \\ &&  
\hbar J \left[ \hat{\psi }^\dag _1(z)\hat{\psi }_2(z)+\hat{\psi }^\dag _2(z)\hat{\psi }_1(z)\right] { \} }\, .
\label{Hmt} 
\end{eqnarray} 
Here $\hbar J$ is the tunnel-coupling matrix element,  $\hat{\cal T}=-[\hbar ^2/(2m)]\partial ^2/\partial z^2-\tilde{\mu }$ with the chemical potential  $\tilde{\mu }=gn_\mathrm{1D}-\hbar J$, and $g=2\hbar \omega _\perp  a_\mathrm{s}$ is the atomic 1D interaction strength in the limit 
$n_\mathrm{1D}a_\mathrm{s}\ll 1$  \cite{Sal}. 

We study this system based on the description of the quasi-condensate properties by a spectrum of Bogoliubov-type modes \cite{MC03}, which are free-particle-like in the short-wavelength limit and phonon-like in the long-wavelength limit \cite{text1}. We model the ``experimental'' realizations of the atomic quasi-condensate fields by implementing a numerical scheme for generating the initial conditions in the truncated Wigner approximation \cite{TW1,TW2} and represent their wave functions as $\psi_j(z) = \sqrt{n_\mathrm{1D}+\delta n_j(z)} \exp[i\phi _j(z)]$, $j=1,2$, where $\phi_j$ and $\delta n_j$ are the local phase and density fluctuations, respectively. We decompose these fluctuations into waves corresponding to elementary excitations of two coupled quasi-condensates by means of an extension of the approach by Mora-Castin \cite{MC03} as developed by Whitlock and Bouchoule \cite{B2}. The amplitudes and node positions of these waves are chosen randomly, assuming the Bose-Einstein statistics of the elementary excitations. In particular, the relative phase $\theta (z)\equiv \phi _1(z)-\phi _2(z)$  is modeled as 
\begin{eqnarray}
\theta (z) &=& \sqrt{2/( n_\mathrm{1D}L_\mathrm{max})}\, \sum _{k\neq 0} 
\left[ ({\eta _k+2\, gn_\mathrm{1D}})/{\eta _k}\right] ^{1/4} \times 
\nonumber \\ && \sqrt{ {\cal B}_k |\ln \xi _k| } \sin (kz+2\pi \xi ^\prime _k) , 
\label{theta}
\end{eqnarray} 
where  $\xi _k$, $\xi ^\prime _k$ are random numbers (obtained by a pseudo-random number generator) uniformly distributed between 0 and 1, $\eta _k= (\hbar k)^2/(2m)+2\hbar J$ and the summation is taken over the discrete spectrum of wave vectors $k$ equal to (both positive and negative) multiples of 
$2\pi /L_\mathrm{max}$ \cite{text1}. A similar expansion holds for the density fluctuations. The explicit dependence of the wave amplitude on the random numbers $\xi_k$  reflects the statistics of the occupation numbers of the elementary-excitation modes. To include both thermal and quantum fluctuations (zero-point oscillations of the atomic field):
\begin{equation} 
{\cal B}_k =2^{-1} \coth [ \, {\varepsilon _k}/({2k_\mathrm{B}T}) \, ] , 
\label{fullB}
\end{equation} 
where $\varepsilon _k =\sqrt{\eta _k(\eta _k+2\, gn_\mathrm{1D})}$ is the energy of the elementary excitation. We refer the use of Eq. (\ref{fullB}) to as the ``full Bogoliubov approach". Alternatively, if we choose to neglect the quantum fluctuations: 
\begin{equation} 
{\cal B}_k \approx {k_\mathrm{B}T}/{\varepsilon _k} .   
\label{excl}
\end{equation} 

We first analyze the full Bogoliubov approach and calculate the full distribution function of the contrast following \cite{A1,A2}: The interference pattern integrated over a sampling length $L$ is characterized by the complex amplitude operator $\hat{A}(L)$ \cite{text4}. Each experimental run yields probabilistically a complex value $A(L)$. The expectation value $\langle \hat{A}(L)\rangle $ is zero, but $\langle \hat{A}(L)^\dag \hat{A}(L)\rangle \equiv \langle | {A}(L)|^2 \rangle  $ is not. It is convenient to study the statistical distribution $W(\alpha)$ where $\alpha \equiv | {A}(L)|^2 / \langle | {A}(L)|^2 \rangle $ is the square of the absolute value of the integrated contrast scaled to its mean.

Using Eq. (\ref{fullB}) we generated integrated contrast distributions for a wide range of parameters.  In Fig.~1 we display $W(\alpha )$ as a function of the sampling length, both for zero and for non-zero coupling. The calculations have been done for $L_\mathrm{max}=100~\mu $m and 60 modes taken into account (increasing of the number of modes to 120 does not change the results significantly).

\begin{figure} 
\epsfig{file=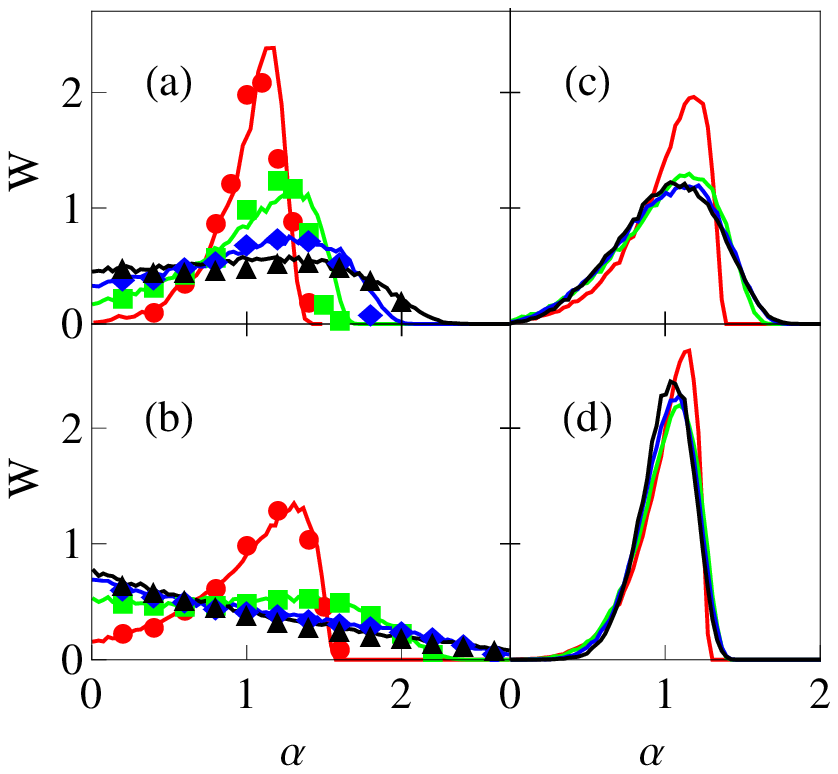,width=0.89 \columnwidth}
\begin{caption}
{(Color online). Interference contrast distribution $W(\alpha )$ of a $^{87}$Rb quasi condensate with $n_\mathrm{1D}=59~\mu $m$^{-1}$ and $\omega _\perp = 2\pi \times 3$~kHz as a function of the sampling length $L$ for 
\emph{(a)}  $T=31 $~nK, $J=0$, \emph{(b)}  $T=60 $~nK, $J=0$, \emph{(c)}  $T=60 $~nK, $J=2\pi \times 1$~Hz, 
and \emph{(d)}   $T=60 $~nK, $J=2\pi \times 3$~Hz. 
Lines represent the results of the full Bogoliubov modeling for $L=10~\mu $m (red), $24~\mu $m (green), $37~\mu $m (blue), and $51~\mu $m (black). 
Symbols (circles, squares, diamonds, up triangles) of the respective colors show the results of the Luttinger liquid approach \cite{tnoise-s} for the same values of $L$ and $J=0$. \label{fig:1} } 
\end{caption}
\end{figure}

In the special case of zero tunnel coupling between the condensates ($J=0$), statistical independence of fluctuations in each quasi-condensate allows to separate correlations: 
$$
\langle | {A}(L)|^2 \rangle = \int _0^L dz\int _0^L dz^\prime \, 
\langle \hat{\psi }_1^\dag (z)\hat{\psi }_1(z^\prime )\rangle \, 
\langle \hat{\psi }_2^\dag (z^\prime )\hat{\psi }_2(z)\rangle ,  
$$
and a general formula for 
the computation of all the moments of $W(\alpha )$ can be found from the Luttinger-liquid formalism \cite{A1,A2}. The stochastic properties of $W(\alpha )$ are then determined by a single dimensionless parameter $\kappa _TL$, where $\kappa _T=mk_\mathrm{B}T/(\hbar ^2 n_\mathrm{1D})$ is the inverse thermal coherence length, $m$ is the mass of the atom. For $^{87}$Rb $\kappa _T\approx 1.815~\mu \mathrm{m}^{-1} (T/100~\mathrm{nK})(10~\mu \mathrm{m}^{-1}/n_\mathrm{1D})$. There is very good agreement between the full Bogoliubov calculations and the Luttinger liquid formalism,  and one observes (Fig.~1a,b) the characteristic change between a Gumbel-like distribution to an exponential distribution as the ratio of the averaging length to the characteristic phase-coherence length grows \cite{tnoise-s}  

If $J\neq0$ and large enough (i.e. $J \sim 2\pi \times 1$~Hz for the typical experimental range of $n_\mathrm{1D}$, $T$, and $L$), we observe a different picture: the distribution  $W(\alpha )$ stabilizes at some peaked shape and preserves this shape as $L$ grows further, (Fig.~1c,d).  This is  characteristic for the \emph{phase locking} between the two matter waves.  Since the Luttinger liquid approach \cite{A1,A2} is based on the assumption of statistical independence of fluctuations in the two quasi-condensates, it can not easily be extended to the tunnel-coupled systems described by Eq. (\ref{Hmt}). 

We can now study the effect of quantum fluctuation by using Eq. (\ref{excl}) instead of Eq. (\ref{fullB}).  For weakly interacting 1D systems the differences are small (Fig.~2). 

This observation suggests a simple semiclassical description of the of the noise properties at distances longer than the healing length $\zeta _\mathrm{h} = {\cal K}/(\pi n_\mathrm{1D})$ (Luttinger parameter ${\cal K}=\pi \hbar \sqrt{n_\mathrm{1D} / (mg)}$), where  density fluctuations are suppressed and the main contribution to noise comes from fluctuations of the relative phase $\theta (z)$. For thermal excitations, the fluctuations of $\theta (z)$ are Gaussian and their autocorrelation 
function is \cite{B2}:
\begin{equation} 
\langle \theta (z) \theta (z^\prime ) \rangle = \kappa _T l_J \exp ( -|z-z^\prime |/l_J), 
\label{autocor} 
\end{equation} 
with $l_J=\frac 12 \sqrt{\hbar /(mJ)}$ . For $^{87}$Rb $l_J \approx 5.367\;\mu \mathrm{m} \, \sqrt{J_1/J}$, where $J_1\equiv 2\pi \times 1\; \mathrm{Hz}$. Eq.~(\ref{autocor}) is valid under the following assumptions: {(i)} we can neglect atom shot noise \cite{text2}, 
{(ii)} the density fluctuations are suppressed, {(iii)} {\em quantum} fluctuations of the phase can be neglected, and the mean occupation number for the given mode is taken in the classical (Boltzmannian) limit Eq. (\ref{excl}). The relative phase evolution along $z$ can be described by an Ornstein-Uhlenbeck stochastic process \cite{sptxt}, where the coordinate $z$ plays the role of time:  
\begin{equation} 
\frac d{dz} \theta (z)= -\frac 1{l_J}\theta (z)+f(z) . 
\label{ousp}
\end{equation}
Here $f(z)$ is the random force with the properties $\langle f(z)\rangle =0$, $\langle f(z_1)f(z_2)\rangle = 2\kappa _T\, \delta (z_1-z_2)$, and $l_J^{-1}$ plays the role of the friction coefficient. The local variance of the relative phase should not depend on $z$ and the initial value $\theta (z^\prime )$ is distributed according to a Gaussian with zero mean and variance $\langle \theta ^2(z^\prime ) \rangle =\kappa _Tl_J$, i.e., the stationary distribution following from Eq. (\ref{ousp}). 

\begin{figure} 
\epsfig{file=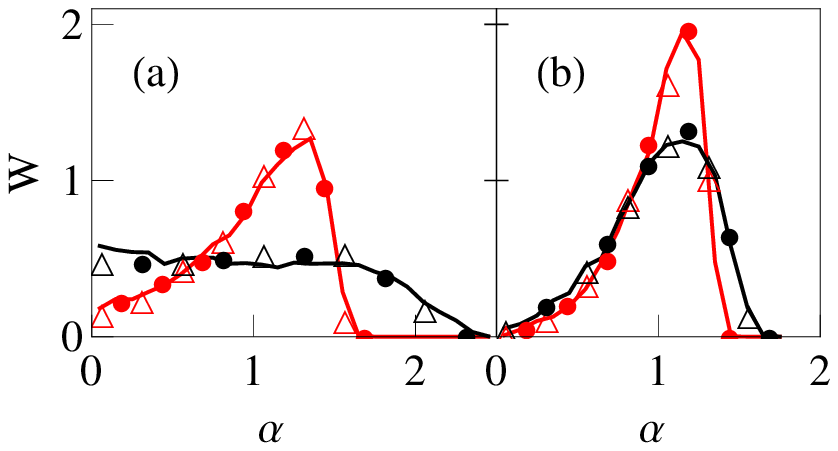,width=0.89 \columnwidth}
\begin{caption} 
{(Color online). Interference contrast distribution 
$W(\alpha ) $ for (a) $\kappa _Tl_J\rightarrow \infty $  and  (b) $\kappa _Tl_J=1.0$. 
Solid lines display the results of the 
Ornstein-Uhlenbeck stochastic process modelling for $\kappa _TL=1.85$ (red) 
and 4.43 (black). Results of 
the full Bogoliubov [Eq.~(\ref{fullB})] simulations and Bogoliubov simulations 
without quantum noise [Eq.~(\ref{excl})] are shown by  triangles and circles, respectively. \label{fig:2} }
\end{caption} 
\end{figure}

This leads to a very simple and efficient way to calculate the fluctuation properties. We propagate $\theta (z)$ from $z=0$ to $z=L$ using an exact updating formula for Eq. (\ref{ousp}) \cite{spsim} and compute for each run the complex phase \cite{text3}. A statistical analysis of the full distribution function of $\alpha $ on the ensemble of runs shows a very good agreement between the Bogoliubov simulations and the stochastic process modeling (Fig. \ref{fig:2}).

\begin{figure}  
\epsfig{file=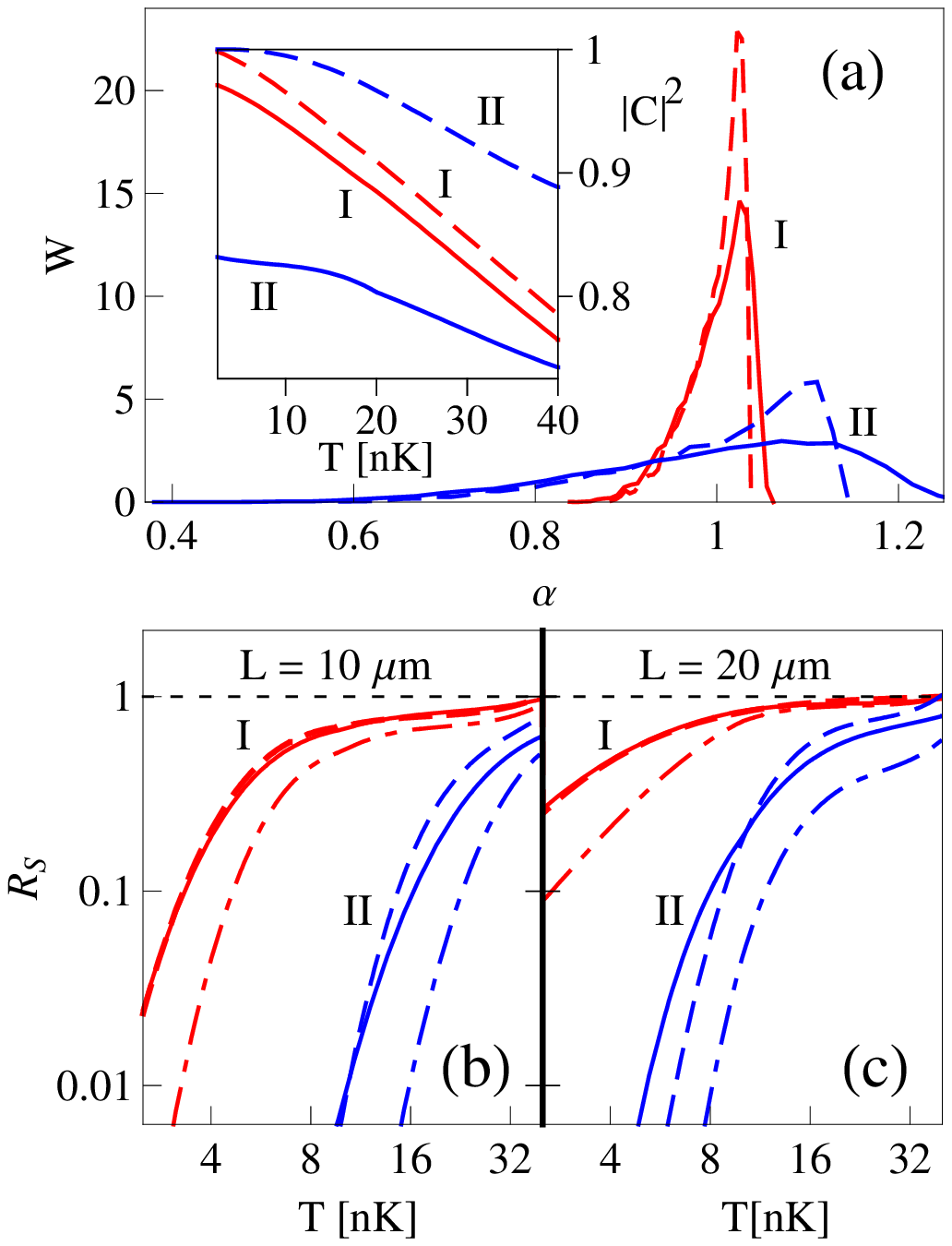,width=0.89 \columnwidth} 
\begin{caption} 
{(Color online). (a) Interference contrast distribution $W(\alpha )$ for $^{87}$Rb atoms,  $n_\mathrm{1D}=60~\mu\mathrm{m}^{-1}$, $J=0$, and $L=10~\mu $m calculated for both thermal and quantum fluctuations (solid line) and thermal fluctuations only (dashed line) for two different parameter sets: $T=10$ nK, $\omega _\perp =2\pi \times 3$ kHz (I, red) and $T=40$ nK, $\omega _\perp =2\pi \times 60$ kHz (II, blue).   
The inset shows the temperature dependence of $|C|^2\equiv \langle |A(L)|^2\rangle /L^2$ (the average square of the dimensionless coherence factor) for the respective cases.
(b) Ratio of the centered $s$th order momenta $R_s$ for $s=2$ (solid line), $s=3$ (dashed line; hardly distinguishable from $s=2$ in the case I) and $s=4$ (dot-dashed line) as a function of temperature for two parameter sets at $L=10~\mu $m. Quantum fluctuations manifest themselves in $R_s < 1$. (c) The same as in (b) but for $L=20~\mu $m.  The effects of atomic shot noise on $\langle |A(L)|^2\rangle $ and $W(\alpha )$ are not taken into account. \label{fig:3} }
\end{caption} 
\end{figure}  

For which parameters and observables are the fundamental quantum fluctuations in 1D systems observable? 

We first analyze the modification of the full distribution function $W(\alpha)$.  The contribution of quantum noise will be detectable in $W(\alpha )$ at very low temperatures $k_\mathrm{B} T \ll \mu $ and short length scales $L\approx 10~\mu $m (Fig.~3).  It can be quantified by the ratio $R_s =\langle (\alpha -1)^s\rangle _\mathrm{th}/\langle (\alpha -1)^s\rangle _\mathrm{q}$ where 
the averages $\langle \dots \rangle _\mathrm{q}$ and $\langle \dots \rangle _\mathrm{th}$  are obtained by either the full Bogoliubov approach including quantum fluctuations [Eq.~(\ref{fullB})] or considering only thermal fluctuations [Eq.~(\ref{excl})], respectively (Fig.~3b,c). If the quantum noise is negligible, then $R_s=1$.  

\begin{figure} 
\epsfig{file=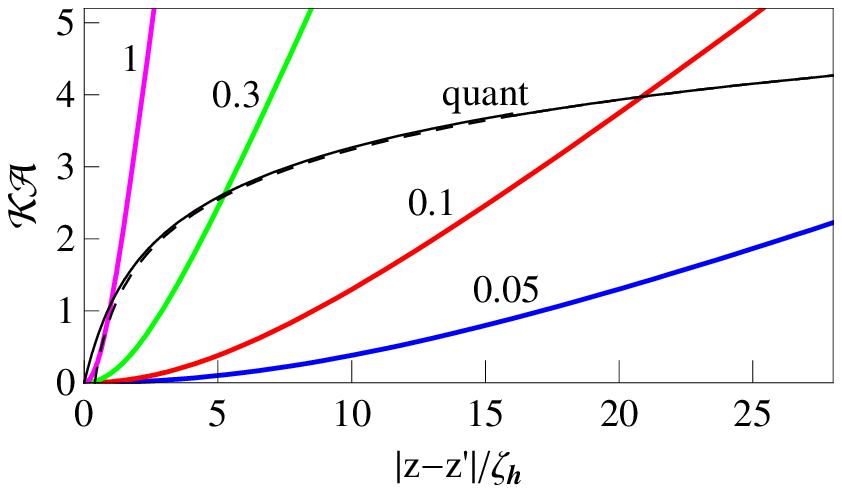,width=0.89 \columnwidth} 
\begin{caption} 
{(Color online). The universal (dimensionless) functions ${\cal KA}_\mathrm{th} $ and 
${\cal KA}_\mathrm{q} $ as a function of distance (in units of healing length) for uncoupled 
quasicondensates. Thin solid line: contribution of the quantum noise; dashed line: its logarithmic 
asymptotics. Thick lines: contributions of the thermal noise; the curves are labelled by the respective 
values of $k_\mathrm{B}T/(gn_\mathrm{1D})=1$, 0,3, 0.1, and 0.05. 
\label{fig:4} } 
\end{caption} 
\end{figure} 

We can quantify the relative contribution of thermal and quantum noise by examining the phase correlation function ${\cal C}_\theta (z-z^\prime) = \left\langle \exp [\mathrm{i}(\theta(z)-\theta(z^\prime))]\right\rangle \equiv \exp [-{\cal A}(|z-z^\prime |)]$. The function ${\cal A}$ that governs the decay of correlations, can be represented as a sum of the quantum (q) and thermal (th) parts: ${\cal A}={\cal A}_\mathrm{q}+ {\cal A}_\mathrm{th}$. In the case of uncoupled quasi-condensates ($J=0$) both have the same dependence on the Luttinger parameter,  they are proportional to ${\cal K}^{-1}$.  Consequently, ${\cal K}{\cal A}_\mathrm{q}$ and ${\cal K}{\cal A}_\mathrm{th}$ are universal functions, depending on $k_\mathrm{B}T/(gn_\mathrm{1D})$ and $|z-z^\prime |/\zeta _\mathrm{h} $ only. In Fig. \ref{fig:4} we plot these functions. 

Analyzing both contributions ${\cal A}_\mathrm{th,q}$ to ${\cal C}_\theta $ we find: At small distances $\partial {\cal A}_\mathrm{th}/\partial |z-z^\prime |\rightarrow 0$; for distances $|z-z^\prime |\gtrsim  \zeta _\mathrm{h}gn_\mathrm{1D}/(k_\mathrm{B}T)$ we recover the linear asymptotics ${\cal A}_\mathrm{th}\approx \kappa _T |z-z^\prime |$.  In contrast ${\cal A}_\mathrm{q}$ is linear in $|z-z^\prime |$ up to the healing length $\zeta _\mathrm{h}$, for larger distances we obtain the asymptotics ${\cal A}_\mathrm{q}\approx \frac 1{\cal K} \ln [ 8|z-z^\prime |/(\pi \zeta _\mathrm{h})] $, i.e.  ${\cal C}_\theta \propto |z-z^\prime |^{-1/{\cal K}}$. 

Consequently thermal fluctuations become dominant at $|z-z^\prime |\gtrsim \zeta _\mathrm{h} gn_\mathrm{1D}/(k_\mathrm{B}T) =\pi /(\kappa _T{\cal K})$. These estimations of the relative contribution of the quantum noise are also valid for coupled systems with $l_J \gg \pi /(\kappa _T{\cal K})$. 

To conclude, we have studied the fluctuation properties in samples of interacting quantum degenerate 1D bosons. In contrast to previous work \cite{tnoise-s,A1,A2}, our approach allows also to investigate also tunnel coupled, phase-locked 1D systems, and provides a clear distinction between contributions of fundamental quantum noise and thermal excitations. In addition we show that on length scales $|z|\gtrsim \pi /(\kappa _T{\cal K})$, where the fluctuations are dominated by thermal excitations, these systems can be described to a very good approximation by a simple semiclassical model based on the spatial evolution of the relative phase according to an Ornstein-Uhlenbeck stochastic process with Gaussian phase fluctuations.  We expect that one can find similar semiclassical models for many other quantum degenerate systems at finite temperature, such as 1D spinor systems \cite{spinor}.

This work was supported by the Austrian Ministry of
 Science via its grant for the WPI, by the WWTF
 (Viennese Science Fund, Project No. MA-45), and by the
 FWF (projects No. Z118-N16 and No. M1016-N16).

 

\begin{thebibliography}{99}
\bibitem{text1} We approximate the trapped cloud by a uniform 1D system and use periodic boundary conditions set on the length $L_\mathrm{max}$, the longitudinal size of the trapped sample (typically $\sim 10^2~\mu $m), which allows to apply the theories developed for uniform systems \cite{Ha81,A1,A2,B2}.
\bibitem{text2} In a strict sense, the correlator in Eq. (\ref{autocor}) assumes normal ordering of operators  and thus eliminates shot-noise effects. In real experimental images \cite{tnoise-s} each individual (local) interference pattern within the optical resolution limit contains more than $400$ atoms, so we can  neglect the shot-noise (about 5 \% uncertainty) and consider Eq. (\ref{autocor}) as a good approximation.
\end{thebibliography}
\end{document}